# How to Manage the Post Pandemic Opening? A Pontryagin Maximum Principle Approach

R. Mansilla[1]

[1] Center for Interdisciplinary Research in Sciences and the Humanities, National Autonomous University of Mexico, Mexico

Correspondence: R. Mansilla, Center for Interdisciplinary Research in Sciences and the Humanities, National Autonomous University of Mexico, Mexico.

**Abstract**

The COVID-19 pandemic has wholly disrupted the operation of our societies. Its elusive transmission process, characterized by an unusually long incubation period and a high contagion capacity, has forced many countries to take quarantine and social isolation measures that conspire against national economies' performance. This situation confronts decision-makers in different countries with the alternative of reopening the economies, thus facing the unpredictable cost of a rebound of the infection. This work tries to offer an initial theoretical framework to handle this alternative.

**Keywords:** dynamic optimization, maximum principle, post pandemic opening

**1. Introduction**

After several months of quarantine and social isolation, the national governments of the different countries face the inevitable decision to reactivate their economies. This decision is accompanied by the almost certain rebound of the epidemic, since, in the absence of social isolation, the number of infections will surely skyrocket again. In the execution of this process, there are two evident extremes:

A restart of economic activities in all their extent will cause a strong rebound of the epidemic with the expected death of many human beings.

Continuing for an extended period of time, quarantine will result in an economic contraction with unpredictable consequences.

Therefore, the restart of the economies and the rebound of the epidemic are in essence, two contradictory forces. The search for a strategy that tries to manage the unwanted effects optimally is necessary.

In the pages that follow, a functional is constructed, subject to certain dynamic restrictions that aim to capture the essential aspects of the process of economic restart under the preexistence of the epidemic. The method used to solve the problem has been the Pontryaguin Maximum Principle. Once the conditions that this principle prescribes have been established, a system of ordinary differential equations (7) and an equation for the control (6) are obtained that allow to maximize the economic performance subject to the restrictions imposed by the contagion. Conditions to avoid the rebound of the epidemic are also analyzed in the text.

**2. Method**

*2.1 Definition of Variables and Problem Statement*

First, the following functions are defined:

$a_\tau(t)$: The number of asymptomatic people in the instant of time t belonging to the age range that makes up the economically active population, but with age less than or equal to τ which are allowed to go out to work. Decision-makers may vary the parameter τ as the epidemiological situation changes. The idea of initially restricting through age to people who are allowed to go out to work is due to the virus's well-known characteristic of being more severe with the elderly. This group contains the susceptible, latent, (that is when the individual has been infected, but does not yet show symptoms or is not still capable of infecting) and asymptomatic infectious, all of them of age less than or equal to $\tau$. Immediately a member of this group is detected as infected, it is quarantined. In all that follows, governments are supposed to be in a position to exercise this measure. Notice that the function $a_\tau(t)$ qualifies as a control function because it has the following two properties. First, it is something that is subject to the discretionary choice of decision-makers. Second, the choice of $a_\tau(t)$ affects the other variables of the problem defined below, as will be seen later.

$q(t)$: People in quarantine at the instant of time $t$. This group includes children, young people younger than the minimum age for working, people older than τ, and anyone who has been detected infected with severity low enough to pass the disease at home. In short, all the people who must stay at home.

$h(t)$: All hospitalized people who are not in intensive care units at the instant of time $t$.

$u(t)$: All people who are in intensive care units in the instant of time $t$.

The problem can be stated as follows:

Find a control function $a_\tau(t)$ that maximizes the functional:

$$\max_{a_\tau} \int_0^T [Aa_\tau^\alpha(t) - \beta_q q(t) - \beta_h h(t) - \beta_u u(t)] e^{-rt} dt \qquad (1)$$

subject to the conditions:

$$\begin{cases} \frac{dq(t)}{dt} = m(q(t), a_\tau(t)) - \nu_{qh} q(t) \\ \frac{dh(t)}{dt} = \nu_{qh} q(t) - \nu_{hu} h(t) \\ \frac{du(t)}{dt} = \nu_{hu} h(t) - \nu_{u\infty} u(t) \end{cases} \qquad (2)$$

where:

$Aa_\tau^\alpha(t)$: It is the aggregate production function [1], which we assume is of the Cobb-Douglas type and depends only on labor. This last assumption is due to the supposition that in the period in which the reopening of the economy is going to take place, the depreciation of physical capital and its variation is negligible. Therefore, the aggregate production function depends only on labor. Constant A contains a factor to transform the people hired into working hours according to the duration of the current working day. Obviously, $0 < \alpha < 1$.

$\beta_q$: The average social cost of one day of quarantine per person. This coefficient considers the loss of classes or the cost of receiving them online, the value of hours of work lost by adults due to childcare, the amount of work lost by quarantined persons who remain at home, etc.

$\beta_h$: Cost of one day of hospitalization per person, for patients not treated in intensive care units.

$\beta_u$: Cost of one day of treatment per person in an intensive care unit.

$r$: Interest rate used to bring the earnings to present value.

$m(q(t), a_\tau(t))$: Rate of increase of $q(t)$ because of the interaction between the infected of the group $a_\tau$ and the quarantines, as well as among the members of the group $a_\tau$. We will assume that this function holds the following conditions:

$$\frac{\partial m}{\partial a_\tau} > 0 \quad ; \quad \frac{\partial^2 m}{\partial a_\tau^2} > 0 \quad ; \quad \frac{\partial m}{\partial q} > 0 \quad ; \quad \frac{\partial^2 m}{\partial a_\tau^2} > 0$$

$\nu_{qh}$: Fraction of quarantines who must be hospitalized.

$\nu_{hu}$: Fraction of hospitalized patients who must be taken to an intensive care unit.

$\nu_{u\infty}$: Fraction of people who are cared for in intensive care units and who leave them for recovery or death. The initial conditions for $q(t)$, $h(t)$, $u(t)$ and $a_\tau(t)$ should be taken the day of the start of the reopening process.

**3. Results.**

*3.1 Pontryaguin Maximum Principle and the Hamiltonian of the problem.*

The Hamiltonian of the problem can be written as follows:

$$\mathcal{H}(a_\tau, q, h, u, \lambda_1, \lambda_2, \lambda_3) = [Aa_\tau^\alpha - \beta_q q - \beta_h h - \beta_u u]e^{-rt} \to$$
$$\to -\lambda_1 [m(q, a_\tau) - \nu_{qh} q] - \lambda_2 [\nu_{qh} q - \nu_{hu} h] - \lambda_3 [\nu_{hu} h - \nu_{u\infty} u]$$

The minus sign in front of the functions $\lambda_i(t)$ it is to emphasize that these shadow prices are costs (see [2], pp. 206-207 or [4], p. 118). Therefore, the functions $\lambda_i(t)$ will be assumed positive.

The formulation of the Pontryaguin Maximum Principle can be seen in many places (see [3], pp. 218-221). Let us now consider the conditions that the functions involved must hold in the case of constituting an extreme.

I) The function $a_\tau(t)$ must be a maximum of the Hamiltonian $\mathcal{H}$:

$$\frac{\partial \mathcal{H}}{\partial a_\tau} = 0 \quad ; \quad \frac{\alpha A}{a_\tau^{1-\alpha}(t)} e^{-rt} - \lambda_1(t) \frac{\partial m(q(t), a_\tau(t))}{\partial a_\tau} = 0 \tag{3}$$

II) If we denote by:

$$\vec{\lambda} = (\lambda_1, \lambda_2, \lambda_3)^T \quad ; \quad \vec{x} = (q, h, u)^T \tag{4}$$

then:

$$\frac{d\vec{\lambda}}{dt} = -\frac{\partial \mathcal{H}}{\partial \vec{x}} \tag{5}$$

*3.2 The Resolution of the Necessary Conditions of Extreme*

Equation (5) is a system of equations that can be written as:

$$\begin{cases} \frac{d\lambda_1}{dt} = \left[\frac{\partial m}{\partial q} - v_{qh}\right]\lambda_1 + v_{qh}\lambda_2 + \beta_q e^{-rt} \\ \frac{d\lambda_2}{dt} = -v_{hu}\lambda_2 + v_{hu}\lambda_3 + \beta_h e^{-rt} \\ \frac{d\lambda_3}{dt} = -v_{u\infty}\lambda_3 + \beta_u e^{-rt} \end{cases}$$

The last two equations of the system are independent of the first and can be solved separately:

$$\frac{d}{dt}\begin{bmatrix}\lambda_2 \\ \lambda_3\end{bmatrix} = \begin{bmatrix}-v_{hu} & v_{hu} \\ 0 & -v_{u\infty}\end{bmatrix}\begin{bmatrix}\lambda_2 \\ \lambda_3\end{bmatrix} + \begin{bmatrix}\beta_h \\ \beta_u\end{bmatrix} e^{-rt}$$

The characteristic polynomial of the system matrix is:

$$p(\omega) = (\omega + v_{hu})(\omega + v_{u\infty})$$

The eigenvalues are:

$$\omega_1 = -v_{hu} \quad ; \quad \omega_2 = -v_{u\infty}$$

and the corresponding eigenvectors:

$$v_1 = \begin{bmatrix}1 \\ 0\end{bmatrix} \quad ; \quad v_2 = \begin{bmatrix}v_{hu} \\ v_{hu} - v_{u\infty}\end{bmatrix}$$

Therefore, component $\lambda_2(t)$ of the general solution of the system can be expressed as follows:

$$\lambda_2(t) = c_1 e^{-v_{hu}t} + c_2 e^{-v_{u\infty}t} + B_0 e^{-rt}$$

where:

$$B_0 = \left[\frac{\beta_h}{v_{hu} - r} + \frac{\beta_u v_{hu}}{v_{hu} - v_{u\infty}}\left(\frac{1}{v_{u\infty} - r} - \frac{1}{v_{hu} - r}\right)\right]$$

Returning to Equation (3):

$$\frac{\alpha A}{\lambda_1(t)} e^{-rt} = \frac{\partial m}{\partial a_\tau} a_\tau^{1-\alpha}$$

Since $\frac{\partial m}{\partial a_\tau} > 0$ ; $\frac{\partial^2 m}{\partial a_\tau^2} > 0$ ; $1 - \alpha > 0$ there is a function $G(x, y)$, invertible with respect to $x$ such that:

$$a_\tau(t) = G\left(\frac{\alpha A}{\lambda_1(t)} e^{-rt}, q(t)\right) \tag{6}$$

Finally, the following system of ordinary differential equations must be solved:

$$\begin{cases} \frac{d\lambda_1(t)}{dt} = \left[\frac{\partial m}{\partial a_\tau}\left(q(t), G\left(\frac{\alpha A}{\lambda_1(t)}, q(t)\right)\right)\right]\lambda(t)_1 + K(t) \\ \frac{dq(t)}{dt} = m\left(q(t), G\left(\frac{\alpha A}{\lambda_1(t)}, q(t)\right)\right) - \nu_{qh}q(t) \\ \frac{dh(t)}{dt} = \nu_{qh}q(t) - \nu_{hu}h(t) \\ \frac{du(t)}{dt} = \nu_{hu}h(t) - \nu_{u\infty}u(t) \end{cases} \quad (7)$$

where:

$$K(t) = \nu_{qh}\left[c_1 e^{-\nu_{hu}t} + c_2 e^{-\nu_{u\infty}t} + \left(B_0 + \frac{\beta_q}{\nu_{qh}}\right)e^{-rt}\right]$$

Note that there are constants $Q, \theta$ such that:

$$|K(t)| \leq Q e^{-\theta t}$$

In this way, the stability and asymptotic stability of the solutions of the system (7) are established from the properties of the associated autonomous system:

$$\begin{cases} \frac{d\lambda_1(t)}{dt} = \left[\frac{\partial m}{\partial a_\tau}\left(q(t), G\left(\frac{\alpha A}{\lambda_1(t)}, q(t)\right)\right)\right]\lambda(t)_1 \\ \frac{dq(t)}{dt} = m\left(q(t), G\left(\frac{\alpha A}{\lambda_1(t)}, q(t)\right)\right) - \nu_{qh}q(t) \\ \frac{dh(t)}{dt} = \nu_{qh}q(t) - \nu_{hu}h(t) \\ \frac{du(t)}{dt} = \nu_{hu}h(t) - \nu_{u\infty}u(t) \end{cases} \quad (8)$$

The solution of the system (7), probably obtained through numerical methods, together with the control function, obtained in Equation (6), should offer decision makers the guideline to follow to reopen the economies.

### 3.3 Sufficient conditions for the maximum.

Sufficient extreme conditions must be established. The Mangasarian Sufficiency Conditions will be used (for details see [4], p.120). To begin, note that the region:

$$\Omega = \{(q, h, u, a_\tau) \in \mathbb{R}^4. q \geq 0, h \geq 0, u \geq 0, a_\tau \geq 0\}$$

where the variables $q, h, u, a_\tau$ are defined is convex. The functions in the second members of the equations of the system (2) are continuous, as well as their first derivatives. Functions under the integral sign within functional (1) also meet those conditions. Therefore, according to the Mangasarian Sufficiency Conditions quoted above it is sufficient verify that the Hamiltonian's Hessian matrix $\mathbb{H}$ is negative semidefinite (see [5], pp. 511-514). The Hessian matrix has the form:

$$\mathbb{H} = \begin{bmatrix} \mathbb{h}_{11} & 0 & 0 & \mathbb{h}_{14} \\ 0 & 0 & 0 & 0 \\ 0 & 0 & 0 & 0 \\ \mathbb{h}_{41} & 0 & 0 & \mathbb{h}_{44} \end{bmatrix}$$

where:

$$\mathbb{h}_{11} = \frac{\partial^2 \mathbb{H}}{\partial q^2} = -\lambda_1 \frac{\partial^2 m}{\partial q^2} \quad ; \quad \mathbb{h}_{14} = \frac{\partial^2 \mathbb{H}}{\partial q \partial a_\tau} = -\lambda_1 \frac{\partial^2 m}{\partial q \partial a_\tau}$$

$$\mathbb{h}_{41} = -\lambda_1 \frac{\partial^2 m}{\partial q \partial a_\tau} \quad ; \quad \mathbb{h}_{44} = \frac{\partial^2 \mathbb{H}}{\partial a_\tau^2} = \frac{\alpha(\alpha - 1)A}{a_\tau^{2-\alpha}} - \lambda_1 \frac{\partial^2 m}{\partial a_\tau^2}$$

The leading principal minors are $\Delta_1 = -\lambda_1 \frac{\partial^2 m}{\partial q^2} < 0, \Delta_2 = 0, \Delta_3 = 0, \Delta_4 = 0$. Therefore, the leading principal minors are alternate in sign so that the odd order ones are less than or equal to zero and the even order ones are greater than or equal to zero. Hence, the Hessian matrix is negative semidefinite and Hamiltonian of the problem is concave.

*3.4 Particular Case for $m(q(t), a_\tau(t))$.*

Suppose that the function $m(q(t), a_\tau(t))$ has the form:

$$m(q(t), a_\tau(t)) = \nu_{qa} q(t) a_\tau(t) + \nu_a a_\tau^2(t)$$

Note that this choice of function $m(q(t), a_\tau(t))$ minimally captures the two possible contagion processes related to the reopening of the economy: the interaction of the members of the quarantined group with those of the group that is allowed to go out to work $\nu_{qa} q(t) a_\tau(t)$ and the contagion among the members of this latter group due to labor relations $\nu_a a_\tau^2(t)$. In this case Equation (3) can be written as:

$$\frac{\alpha A}{\lambda_1(t)} e^{-rt} = \nu_{qa} q(t) a_\tau^{1-\alpha}(t) + 2\nu_a a_\tau^{2-\alpha}(t)$$

From which can be obtained $a_\tau(t)$.

In the following pages, a qualitative approach to the phenomenon of epidemic rebound is discussed.

*3.5 Possible Scenarios for the Rebound of the Epidemic.*

The simplest procedure to control the rebound of the epidemic is to monitor the values of the function:

$$d(t) = m\left(q(t), G\left(\frac{\alpha A}{\lambda_1(t)}, q(t)\right)\right) - \nu_{qh} q(t)$$

calculated on the path obtained through equation (6) and the system (7). Note that $d(t)$ is the second member of the differential equation in the system (2) corresponding to quarantines. As long as $d(t) < 0$ the number of people in quarantine will be decreasing. But every time $d(t) \to 0^-$, this should be an early warning to decision makers that a new rebound of epidemic is coming.

One of the possible causes of a rebound of the epidemic is the existence of isolated quasi-periodic or periodic solutions asymptotically stable of the system (8). Under the hypothesis that system (8) has periodic solutions, a condition at least necessary to avoid epidemic rebound is that one of the Lyapunov exponents of the solutions mentioned above be positive. The Lyapunov exponents must be calculated numerically in most cases, and the algorithms are generally large CPU time consumers.

**4. Conclusions**

In this work, a minimal theoretical framework has been developed to guide decision-makers from different governments in a reopening of the economy. Equation (6) and system (7) provides the necessary information for this. Functional (1) captures the essential ingredients of this process. Obviously, a refinement of it and the system (2) could provide more precise results.

**Acknowledgements**

The author wishes to thank the warm hospitality received at the Peninsular Center for Humanities and Social Sciences in Mérida, Yucatan, where much of the work presented was carried out.

**Copyrights**